\documentclass[aps,twocolumn,superscriptaddress]{revtex4-1}
\usepackage{}
\usepackage{amssymb}
\usepackage{amsfonts}
\usepackage{graphics,graphicx,epsfig,bm,amsmath,amsthm,amssymb}
\usepackage{bm}
\usepackage{bbm}
\usepackage{longtable}
\usepackage{multirow}
\usepackage{array}
\usepackage{color}
\usepackage[usenames,dvipsnames]{xcolor}

\usepackage{float}
\usepackage[a4paper,colorlinks=true,
linkcolor=blue,citecolor=blue,
pdfauthor={ },
pdftitle={ },
pdfsubject={ },
pdfkeywords={ }]{hyperref}

\begin{document}

\title{
	Perfect coherent transfer in an on-chip reconfigurable nanoelectromechanical network
}

\author{Tian Tian}
\affiliation{CAS Key Laboratory of Microscale Magnetic Resonance and Department of Modern Physics, University of Science and Technology of China, Hefei 230026, China}
\affiliation{Hefei National Laboratory for Physical Sciences at the Microscale, University of Science and Technology of China, Hefei 230026, China}
\affiliation{Synergetic Innovation Center of Quantum Information and Quantum Physics, University of Science and Technology of China, Hefei 230026, China}

\author{Shaochun Lin}
\affiliation{CAS Key Laboratory of Microscale Magnetic Resonance and Department of Modern Physics, University of Science and Technology of China, Hefei 230026, China}
\affiliation{Hefei National Laboratory for Physical Sciences at the Microscale, University of Science and Technology of China, Hefei 230026, China}
\affiliation{Synergetic Innovation Center of Quantum Information and Quantum Physics, University of Science and Technology of China, Hefei 230026, China}

\author{Liang Zhang}
\affiliation{CAS Key Laboratory of Microscale Magnetic Resonance and Department of Modern Physics, University of Science and Technology of China, Hefei 230026, China}
\affiliation{Hefei National Laboratory for Physical Sciences at the Microscale, University of Science and Technology of China, Hefei 230026, China}
\affiliation{Synergetic Innovation Center of Quantum Information and Quantum Physics, University of Science and Technology of China, Hefei 230026, China}

\author{Peiran Yin}
\affiliation{CAS Key Laboratory of Microscale Magnetic Resonance and Department of Modern Physics, University of Science and Technology of China, Hefei 230026, China}
\affiliation{Hefei National Laboratory for Physical Sciences at the Microscale, University of Science and Technology of China, Hefei 230026, China}
\affiliation{Synergetic Innovation Center of Quantum Information and Quantum Physics, University of Science and Technology of China, Hefei 230026, China}

\author{Pu Huang}
\affiliation{National Laboratory of Solid State Microstructures and Department of Physics, Nanjing University, Nanjing 210093, China}

\author{Changkui Duan}
\affiliation{CAS Key Laboratory of Microscale Magnetic Resonance and Department of Modern Physics, University of Science and Technology of China, Hefei 230026, China}
\affiliation{Hefei National Laboratory for Physical Sciences at the Microscale, University of Science and Technology of China, Hefei 230026, China}
\affiliation{Synergetic Innovation Center of Quantum Information and Quantum Physics, University of Science and Technology of China, Hefei 230026, China}

\author{Liang Jiang}
\affiliation{Pritzker School of Molecular Engineering, University of Chicago, Chicago, Illinois 60637, USA}

\author{Jiangfeng Du}
\email{djf@ustc.edu.cn}
\affiliation{CAS Key Laboratory of Microscale Magnetic Resonance and Department of Modern Physics, University of Science and Technology of China, Hefei 230026, China}
\affiliation{Hefei National Laboratory for Physical Sciences at the Microscale, University of Science and Technology of China, Hefei 230026, China}
\affiliation{Synergetic Innovation Center of Quantum Information and Quantum Physics, University of Science and Technology of China, Hefei 230026, China}



\begin{abstract}
	Realizing a controllable network with multiple degrees of interaction is a challenge to physics and engineering.
	Here, we experimentally report an on-chip reconfigurable network based on nanoelectromechanical resonators with nearest-neighbor (NN) and next-nearest-neighbor (NNN) strong couplings. By applying different parametric voltages on the same on-chip device, we carry out perfect coherent transfer in NN and NNN coupled array networks. Moreover, the low-loss resonators ensure the desired evolution to achieve perfect transfer and the demonstration of the parity-dependent phase relation at transmission cycles. The realization of NNN couplings demonstrates the capability of engineering coherent coupling beyond a simple model of a NN coupled array of doubly clamped resonators. Our reconfigurable nanoelectromechanical network provides a highly tunable physical platform and offers the possibilities of investigating various interesting phenomena, such as topological transport, synchronization of networks, as well as metamaterials.	
	
\end{abstract}
\maketitle

\section{Introduction}
An oscillator network, made up of multiple individual resonators and couplings between these nodes, has huge potential in investigating collective phenomena, such as exotic states~\cite{Matheny2019}, symmetry breaking~\cite{Molnar2020}, chimera states~\cite{Hart2016}, an Ising machine~\cite{Marandi2014}, and synchronization~\cite{Holmes2012}.
Recently, a nanoelectromechanical system was proposed and utilized to explore oscillator networks and the associated phenomena~\cite{Fon2017,Matheny2019}.
For an ideal network, one of the most important targets is increasing the number of individual nodes.
Although there are plenty of unit cells in some reported nanomechanical systems~\cite{Awad2017,Cha2018, Cha2018a,Hatanaka2014}, any cell of these networks cannot be individually tuned once the fabrication is completed, which unfortunately limits the realization of arbitrarily configurable graphs and further integration.

Another challenge is the tunability of individual couplings in a large oscillator network. For example, there are some studies that investigated the quantum dot or the spin in a mechanical resonator~\cite{Bennett2013,Yeo2013,Montinaro2014,Ovartchaiyapong2014,Teissier2014,Lee2016,Munsch2017,Carter2018} and explored the possibility of extending the resonators to a large network~\cite{Rabl2010,Machielse2019,Doster2019}. However, so far, strong coupling was demonstrated between only two nanomechanical pillars~\cite{Doster2019}. Since the coupling method between nanomechanical resonators is based on the strain distribution, it is evident that the individual tunability of manipulation will decrease as the number of resonators in networks increases. Therefore, realizing a reconfigurable nanomechanical network with excellent individual tunability is necessary, but it has thus far remained elusive.

Built upon our previous works~\cite{Tian2019,zhang2020coherent}, here we successfully extend the parametric coupling from one-dimensional nearest-neighbor (NN) high-quality-factor resonators to next-nearest-neighbor (NNN) ones, and we put forward an on-chip reconfigurable nanoelectromechanical network. By controlling external voltages, we first show that the couplings of NN and NNN resonators can be independently changed, and the strong-coupling regime can be reached. Then we demonstrate perfect coherent transfer in NN as well as NNN coupled networks in the same on-chip device. The state of each resonator can be measured by lock-in amplifiers. The experiment unambiguously shows that the input excitation is perfectly transferred to the target resonator at transmission cycles, which is consistent with the theoretical prediction. Finally, we verify the phase coherence of the perfect coherent transfer in different array networks.

\section{Perfect-transfer scheme}

The perfect-transfer scheme was proposed in a spin network and a harmonic oscillator’s chain for transferring a quantum state~\cite{Christandl2004,Plenio2004,Yao2013a}. In the protocol, the Hamiltonian is

\begin{figure}\centering
	\includegraphics[width=1\columnwidth]{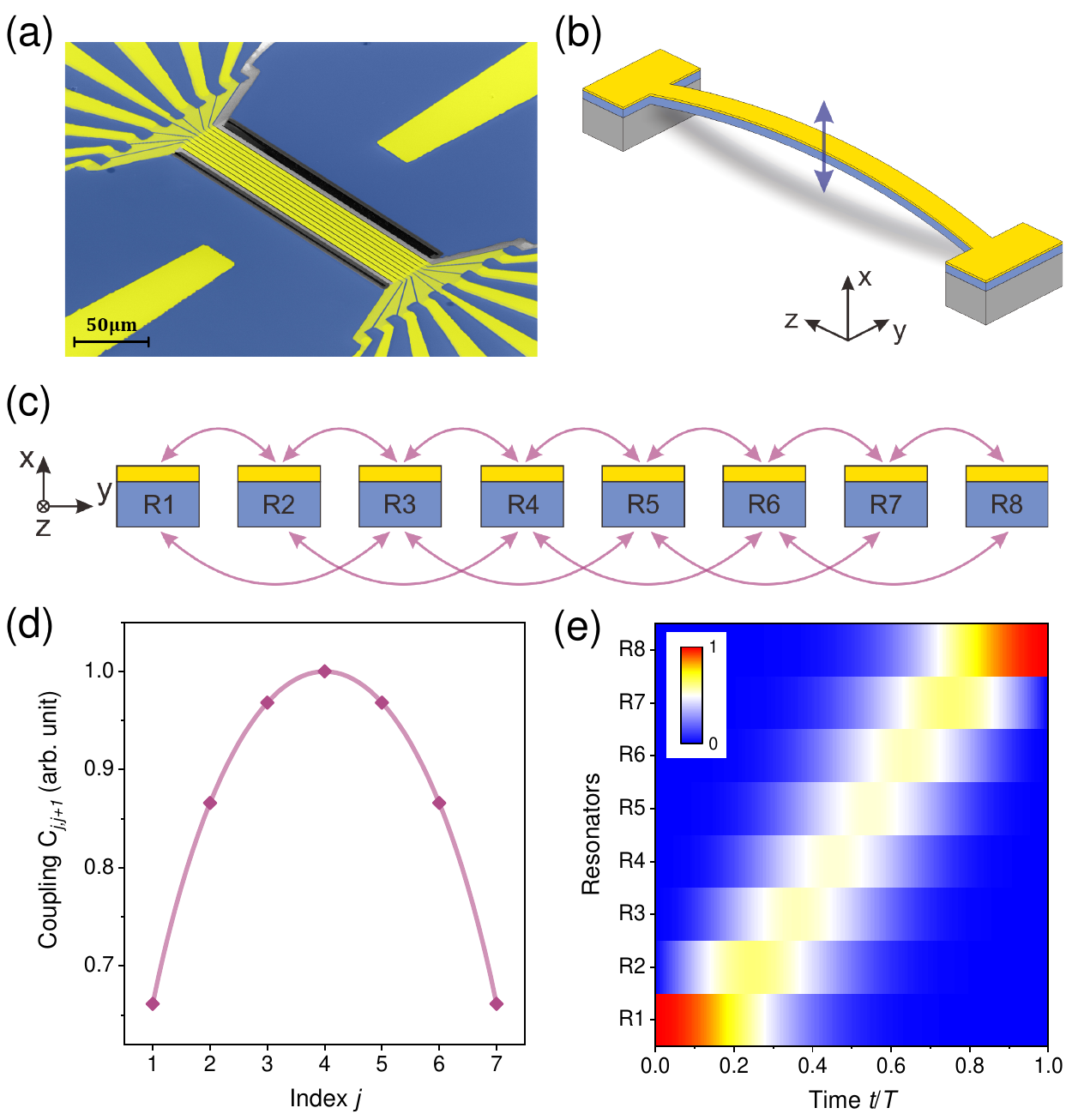}
	\caption{On-chip reconfigurable nanoelectromechanical network and perfect transfer scheme.
		(a) False-color scanning electron micrography of the on-chip device, which includes eight nearly identical nanomechanical resonators. The space between adjacent doubly clamped resonators is 500 nm. (b) Sketch map of the geometric structure and the fundamental out-of-plane mode of a 200-$\mu$m-long, 3-$\mu$m-wide, and 100-nm-thick silicon nitride resonator. These resonators are coated with a 10 nm thin layer of gold to enable electrical manipulation. The magnetic field is applied along the $y$ axis for excitation and detection. (c) The reconfigurable nanoelectromechanical network based on eight resonators with NN and NNN couplings. (d) The couplings of perfect transfer fulfill the relationship $C_{j,j+1} \propto \sqrt{j(N-j)}$ in an array. (e) Perfect coherent transfer from the first site. The color bar stands for excitation amplitudes.
	}
	\label{Set-up}
\end{figure}

\begin{equation}
H=\sum_{j}^{N-1}C_{j,j+1}[a_{j}^{\dagger} a_{j+1} + a_{j+1}^{\dagger} a_{j}],
\label{XY model}
\end{equation}
where $a_{j}^{\dagger}$ and $a_{j}$ are the creation and annihilation operators associated with the $j$th oscillator mode. The coupling $C_{j,j+1}$ describe the interaction between $j$th and $(j+1)$th mode. As shown in Fig.~\ref{Set-up}(d), when the couplings $C_{j,j+1}$ satisfy the mirror-periodic condition
\begin{equation}
C_{j,j+1}=\frac{C_0}{2}\sqrt{j(N-j)},
\label{PST condition}
\end{equation}
a quantum state of the $j$th qubit can be perfectly transferred to the $(N-j+1)$th qubit after a period $T=\pi / C_0$, where $C_0$ is the characteristic coupling strength.

To date, the transfer protocol has been demonstrated in nuclear spins~\cite{Zhang2005}, optical waveguides~\cite{Bellec2012,Perez-Leija2013,Chapman2016}, and superconducting qubits~\cite{Li2018}.
Actually, the core of this protocol ensures the	excited part in the single-excitation subspace, and there is a phase factor $(-1)^{N-1}$ accumulated after a full cycle of forward and backward transfer that only depends on the parity of the number of nodes participating in the transfer protocol~\cite{Christandl2004}. However, the parity-dependent phase relation has not been demonstrated experimentally so far.

\section{on-chip reconfigurable nanoelectromechanical network}

In Fig.~\ref{Set-up}(a), the on-chip nanomechanical system is fabricated by high-stress (1 GPa) silicon nitride, as is detailed in Ref.~\cite{Tian2019}. Each unit cell of this network is a doubly clamped resonator that is 200 $\mu$m long, 3 $\mu$m wide, and 100 nm thick. The spacing of NN resonators is about 500 nm. The eight resonators are nearly identical, except for a small difference in the frequencies caused by the variation in length.
Actually, we take advantage of this small difference in the first vibration frequencies to realize parametric couplings between adjacent resonators~\cite{Huang2013}. As shown by the blue arrow in Fig.~\ref{Set-up}(b), the fundamental mode used is out-of-plane along the $x$ axis for each resonator.
To decrease the dissipation and stable frequencies of these nanoelectromechanical resonators, we put this chip in a vacuum chamber of $1.2 \times 10^{-8}$ mbar and cool it to liquid-nitrogen temperature (77 K). Under this condition, the frequency ranges from 860 kHz to 902 kHz, and quality factors $Q$ reach about $1 \times 10^{5}$. The detailed parameters of eight resonators are listed in Supplemental Material~\cite{Supplement}.

In Fig.~\ref{Set-up}(a) and the cross-section of resonators in Fig.~\ref{Set-up}(c), the thin layer of gold (golden yellow) on the resonators plays two important roles. First, the layer of gold is an electric conductor for each resonator and it converts the radio frequency voltages to mechanical oscillation and vice versa, which ensures excitation and measurement in the magneto-motive technique~\cite{Cleland1999}. Second, it provides capacitive interaction with the adjacent resonators and generates excitation hopping between them, assisted by the parametric drive~\cite{Huang2013}.

As shown in Fig.~\ref{Set-up}(c), all of the nodes (resonators) can be linked by the NN as well as NNN parametric interactions. In this on-chip nanoelectromechanical network, each node can also be controlled and read out by external circuits. We demonstrate this by realizing the linearly tunable couplings and observing perfect coherent transfer in different networks.

\subsection{Nearest-Neighborhood Couplings}

As shown in Fig.~\ref{NN coupled}(a), applying voltages $V_{\rm dc}$ together with $V_{\rm ac}$ in one of the NN resonators leads to the coupling between them. Considering the first bias-tee in Fig.~\ref{NN coupled}(a), here the voltage $V_{\rm ac}$ is the sum of $V_{12}\cos(\omega^{\rm p}_{12}t)$ with $\omega^{\rm p}_{12}=\omega_{1}-\omega_{2}$ and $V_{23}\cos(\omega^{\rm p}_{23}t)$ with $\omega^{\rm p}_{23}=\omega_{2}-\omega_{3}$. These voltages $V_{12}$ and $V_{23}$ are used to control the NN parametric couplings $C_{12}$ and $C_{23}$, respectively.

The experimental result of typical NN parametric coupling is shown in Fig.~\ref{NN coupled}(b), which is measured from the frequency response of the first resonator $R1$. In the measurement, all voltages $V_{\rm dc}$ are chosen as 4 V to keep the frequency difference $\omega^{\rm p}_{12}$ unchanged. When $V_{12}$ = 0 V, the frequency response suggests the frequency $\omega_{1}/2\pi $ = 884.951 kHz and damping rate $\gamma_1 /2\pi$ = 8.17 Hz for the first resonator. It is clear that the coupling strength increases linearly with the voltage $V_{12}$ in Fig.~\ref{NN coupled}(b). The coupling reaches about the linewidth when $V_{12}$ = 0.05 V and about 18$\gamma_1$ when $V_{12}$ = 0.8 V~\cite{Supplement}. The strong coupling region ($V_{12} > $  0.05 V) is indicated by the white line in Fig.~\ref{NN coupled}(b). Based on this couplings, we can implement an array network with eight coupled resonators.

\subsection{Next-Nearest-Neighborhood Couplings}

We extend the NN parametric couplings to NNN resonators in the same device. As described in Fig.~\ref{NNN coupled}(a), we realize the coupling between the $(j-1)$th and $(j+1)$th resonators by applying voltages in $j$th resonator. We show a typical NNN coupling in Fig.~\ref{NNN coupled}(b), where $V_{\rm dc}$ and $V_{57}\cos[(\omega_{5}-\omega_{7})t]$ are combined by a bias-tee and applied to $R6$. Because the capacitance of NNN resonators is smaller than NN ones, we set $V_{\rm dc}=$ 5.5 V to increase the electrostatic force. This typical NNN coupling is measured at the seventh resonator $R7$. Similar to the NN coupling, the NNN coupling strength also increases linearly with the voltage $V_{57}$, and it can reach the strong coupling region when $V_{57} > $  0.05 V. For example, the strength reaches about 18$\gamma_7$ when $V_{57}$ = 0.8 V. The realization of NNN couplings makes it possible to implement an array network with NNN coupled resonators.

\section{Equation of motion}

The motion of $j$th resonator in the one-dimensional coupled nanoelectromechanical array is described by
\begin{equation}
\ddot{x}_j+\gamma_j \dot{x}_j+\omega_{j}^2 x_j=P_{j,j+1}(t)(x_{j+1}-x_{j})+P_{j-1,j}(t)(x_{j-1}-x_{j}),
\label{mechanical equation}
\end{equation}
where $\omega_j$ and $\gamma_j$ is the eigenfrequency and the damping rate of $j$-th resonator. $P_{j,j+1}(t)=C_{j,j+1}\cos(\omega_{j,j+1}^{\rm p} t)/m$ is the dynamic parametric field with coupling strength $C_{j,j+1}$ between the $j$th and $(j+1)$th resonators. According to the electrostatic force parametric coupling, the coupling $C_{j,j+1}$ is proportional to the product of $V_{\rm dc}$ and $V_{j,j+1}$~\cite{Huang2013}.

\begin{figure}
	\includegraphics[width=1\columnwidth]{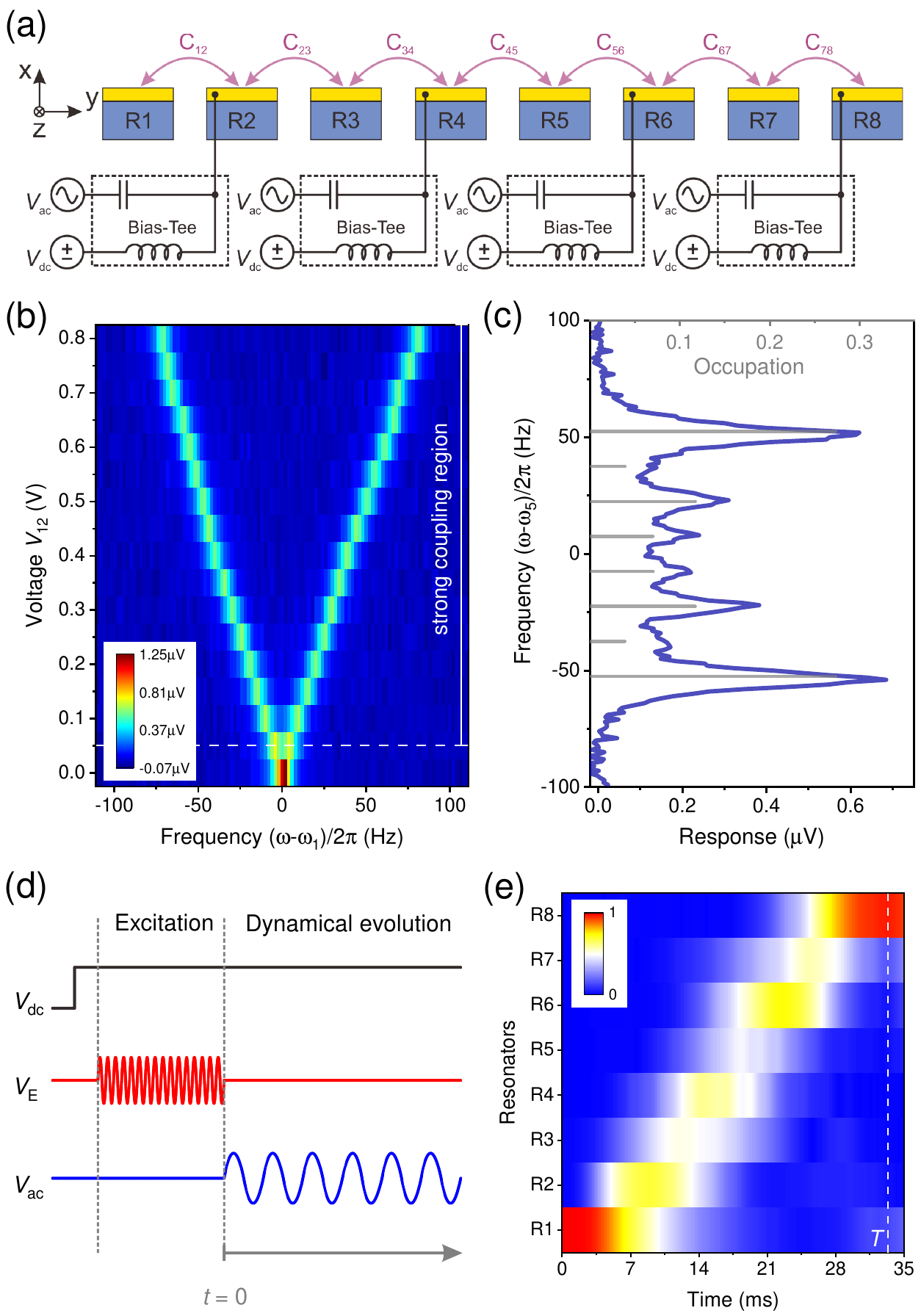}
	\caption{Perfect coherent transfer in the nearest-neighbor coupled nanomechanical network.
		(a) Equivalent circuit diagram of controlling NN parametric couplings. The eight resonators are labeled from $R1$ to $R8$ along the $y$ axis. The dc voltage $V_{\rm dc}$ is combined with an ac voltage $V_{\rm ac}$ by a bias tee, where $V_{\rm ac}$ is used to control neighboring couplings $C_{j,j+1}$. The other end of the resonator (along $z$ axis) is in series with a high resistance (1 M$\Omega$) to make the entire resonator at high potential. (b) Typical parametric coupling between two adjacent resonators under $V_{\rm dc}$ = 4 V. It is measured from the first resonator. The coupling strength increases linearly with the ac voltage $V_{\rm ac}$. (c) The frequency response (dark blue line) at fifth resonator $R5$ under the eight coupled resonators. The gray lines give the theoretical prediction. (d) The pulse sequence in the experiment.
		(e) The experimental result of the perfect coherent transfer from the first resonator. The color bar denotes normalized amplitudes. The dashed white line marks the moment $T$ = 33.3 ms.
	}
	\label{NN coupled}
\end{figure}

By using the complex amplitude $X_{j}(t)$ to characterize the dynamics of  $x_j=\Re(X_je^{i\omega_j t})$ in equation \eqref{mechanical equation}, we can simplify the dynamical evolution of mechanical oscillation~\cite{Tian2019}. Under the rotating wave approximation, $X_{j}(t)$ obeys the following equation:
\begin{equation}
2i \frac{d}{d t} X_{j}(t)=C_{j-1,j} X_{j-1}(t) + C_{j,j+1} X_{j+1}(t).
\label{quantum equation}
\end{equation}
Here, the coupling strength $C_{j,j+1}$ is described by angular frequency, and $X_{j}$ denotes the complex amplitude of $j$th resonator. Because all damping of resonators is nearly identical (i.e. $\gamma_1 \approx \gamma_2 \approx...= \gamma$), it only introduces an overall scaling of $\exp[-\gamma t/2]$ of the amplitudes. We can directly compare the normalized amplitudes $X=(X_1, X_2, X_3, ..., X_8)^T$ at every moment in experiment with theory. Therefore, we map the dynamical evolution of coupled nanoelectromechanical resonators to a Schr\"{o}dinger-like equation, which describes the single excitation tight-binding model in a one-dimensional lattice. For perfect transfer, the period becomes $T=2\pi / C_0$ because of the factor in equation \eqref{quantum equation}.

\section{Perfect coherent transfer in a nearest-neighbor coupled nanomechanical network}

	In experiment, by tuning the ac voltage, the NN couplings $C_{j,j+1}$ can satisfy the condition $C_{j,j+1}=C_{0} \sqrt{j(N-j)}/2$ in Fig.~\ref{Set-up}(d). Specifically, we choose $C_0$ = $2\pi \times 30$ Hz and $N=8$ in this work.
While applying the seven individual NN couplings, we measure the global frequency response at $R5$ to make sure the particular network was realized. As shown in Fig.~\ref{NN coupled}(c), the measured profile of frequency response (dark blue line) is in good agreement with the equidistant theoretical prediction (gray bars), which is crucial for the perfect excitation transfer~\cite{Christandl2004,Plenio2004,Yao2013a}.

Fig.~\ref{NN coupled}(d) shows the pulse sequences for measuring excitation transfer. In the whole experimental procedure, all $V_{\rm dc}$ (black line) are high enough to ensure that their frequencies are stable. First, the resonator $R1$ is excited via applying a radio frequency pulse signal $V_{\rm E}$ (red line). Then, we switch off the excitation source and turn on all $V_{\rm ac}$, which induces excitation transfer across the eight coupled resonators.
The measurement is divided into two parts, specifically. The amplitudes of odd resonators are measured when applying voltages into even ones. Similarly, the amplitudes of even resonators can be measured when applying voltages into odd ones~\cite{Tian2019}.

\begin{figure}
	\includegraphics[width=1\columnwidth]{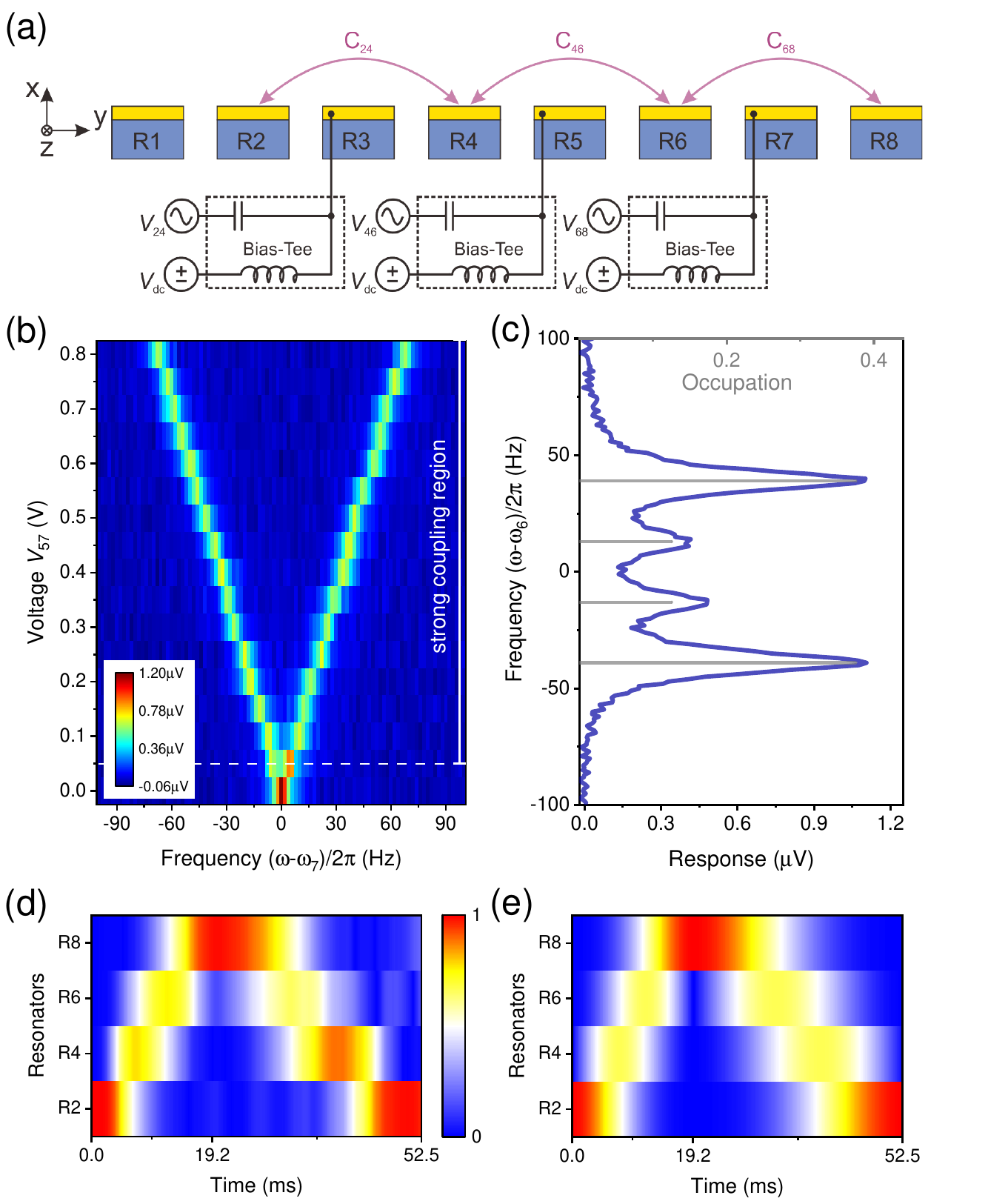}
	\caption{Perfect coherent transfer in the next-nearest-neighbor coupled nanomechanical network.
		(a) Equivalent circuit diagram of controlling NNN parametric couplings. The ac voltages $V_{24}$, $V_{46}$ and $V_{68}$ are used to control $R2-R4$, $R4-R6$ and $R6-R8$ couplings respectively. (b) Typical NNN coupling between $R5$ and $R7$ under $V_{\rm dc}$ = 5.5 V. The measurement from seventh resonator. (c) The frequency response (dark blue line) of sixth resonator $R6$ under the four coupled resonators $R2-R4-R6-R8$ with $C_{0}$ = $2\pi \times 52$ Hz. The gray bars show the theoretical prediction.  Experiment (d) and simulation (e) of excitation evolution in the time-dependent network. The first period $T_{1}$ = 19.2 ms and the second period $T_{2}$ = 33.3 ms represent coupling $C_{0}$ = $2\pi \times 52$ Hz and $2\pi \times 30$ Hz, respectively.
	}
	\label{NNN coupled}
\end{figure}

For each part of dynamical evolution, the amplitudes are demodulated by standard lock-in amplifiers at fixed frequencies of these resonators. To make the results distinct in the whole time domain, we normalize all the amplitudes at each moment; see Fig.~\ref{NN coupled}(e). Obviously, the initial excitation from $R1$ perfectly transfers to $R8$ at the moment $T$ = 33.3 ms. Comparing with Fig.~\ref{Set-up}(e), the experimental result is in agreement with numerical simulation. The minor difference between Fig.~\ref{NN coupled}(e) and Fig.~\ref{Set-up}(e) is caused by ignoring the difference in decay rates of resonators. The original oscillation amplitudes in this process also suggest perfect excitation transfer~\cite{Supplement}.

\section{Perfect coherent transfer in a next-nearest-neighbor coupled nanomechanical network}
According to the circuit diagram shown in Fig.~\ref{NNN coupled}(a), we apply voltages in $R3$, $R5$, and $R7$ to configure one-dimensional NNN coupled network $R2-R4-R6-R8$.
To meet the condition of Eq.\eqref{PST condition} for a chain of 4 resonators, we set $C_{24}$ = $2\pi \times 45$ Hz, $C_{46}$ = $2\pi \times 52$ Hz and $C_{68}$ = $2\pi \times 45$ Hz by tuning ac voltages $V_{24}$, $V_{46}$ and $V_{68}$ respectively. This is in accordance with the situation of $C_{0}$ = $2\pi \times 52$ Hz.
To confirm the realization of a NNN coupled network, one can check the frequency response of any resonator in the experiment. Fig.~\ref{NNN coupled}(c) plots the frequency response, which is measured at the sixth resonator $R6$. The dark blue line shows four distinct peaks, in agreement with the equidistant theoretical prediction labeled by gray bars.

We also carry out perfect coherent transfer in the real-time reconfigurable NNN coupled nanomechanical network.
More specifically, we configure two structures in time domain, namely the coupled structure with $C_{0}$ = $2\pi \times 52$ Hz for a one-transmission cycle $T_{1}$, and then another structure with $C_{0}$ = $2\pi \times 30$ Hz for $T_{2}$. To complete the initial excitation in the edge site, we apply a radio frequency pulse signal $V_{\rm E}$ at $R2$. After that, all NNN couplings are established by turning on the time-dependent voltages $V_{\rm ac}$. We then measure the dynamical evolution.
Lock-in amplifiers demodulate the real-time oscillation amplitudes of $R2$, $R4$, $R6$, and $R8$.
As shown in Fig.~\ref{NNN coupled}(d), the behavior of excitation evolution meets the perfect transfer scheme. It is clear that the excitation initially from $R2$ is transferred to $R8$ at time $T_{1}$ = 19.2 ms, and returns to $R2$ at $T_{1}+T_{2}$ = 52.5 ms. The experimental result is in agreement with the numerical simulation (as shown in Fig.~\ref{NNN coupled}(e)).

\section{Verification of the parity-dependent phase relation}

\begin{figure}
	\includegraphics[width=1\columnwidth]{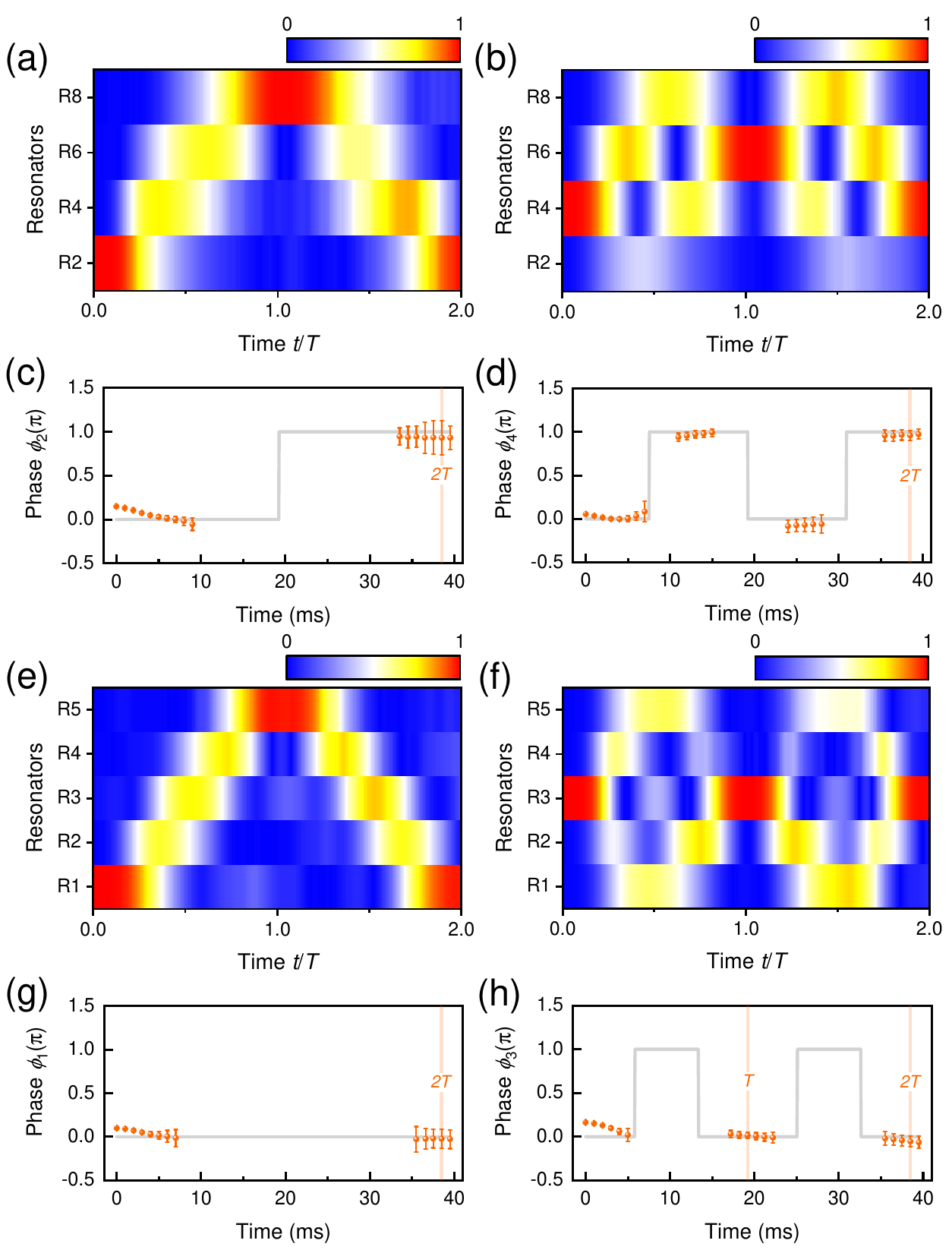}
	\caption{Dynamics of perfect coherent transfer for the NNN coupled network with $N$ = 4 (a-d) and the NN coupled network with $N$ = 5 (e-h).
		The evolution of amplitude distribution over the relevant oscillator nodes (a,b,e,f) and the evolution of phase at the excitation launching site (c,d,g,h), with different choices of launching sites. At time $2T$, we observe a $\pi$ phase shift for oscillator network with even $N$ = 4 (c,d), while no phase shift for odd $N$ = 5 (g,h), regardless of the choice of the launching site.
		Error bars show one standard deviations for 500 repetitions of the experiment. All gray lines are numerical results of quantum mechanics. We choose the same $C_{0}$ = $2\pi \times 52$ Hz for the two networks. The light orange lines stand for the moment $t=T$ and $t=2T$.
	}
	\label{experiment-PET}
\end{figure}

In Fig.~\ref{experiment-PET}, we experimentally verify the phase coherence associated with the perfect transfer from different initial excitation. After a full cycle of forward and backward transfer, the excitation will reappear at the launching site, with a phase change that depends on the number of oscillators, $N$. In Fig.~\ref{experiment-PET}(a,b), we measure the dynamical evolution for the input excitation at $R2$ ($R4$) in the NNN coupled network with $N$ = 4 resonators.
At time $T$, the forward transfer is completed, with excitations transfered to the targeted mirror-symmetric site $R8$ ($R6$).
At time $2T$, a full cycle of forward and backward transfer is completed, with all excitations reappearing at the launching site. Moreover, we also measured a $\pi$ phase shift for oscillator network with $N=4$ in Fig.~\ref{experiment-PET}(c,d), regardless of the launching site of the excitation, which is consistent with the theoretical prediction of phase accumulation of $(-1)^{N-1}$~\cite{Christandl2004}.	

Fig.~\ref{experiment-PET}(e) and Fig.~\ref{experiment-PET}(f) show the perfect coherent transfer for initial excitation at $R1$ and $R3$ in the NN coupled network with $N$ = 5 resonators, respectively.
At time $2T$, there is no phase shift when excitations reappeared at the launching site, consistent with the theory prediction of phase change of $(-1)^{N-1}$ that is trivial for odd $N$.
Thanks to the fact that the target resonator is the same as the initially excited one in Fig.~\ref{experiment-PET}(f), we can further confirm that the phase variation is zero at time $T$ from Fig.~\ref{experiment-PET}(h).
Because the initial demodulation of lock-in amplifiers leads to unreliable data, there are deviations at initial milliseconds in Figs.~\ref{experiment-PET}(c), (d), (g), and (h). Other than this, the experimental data are consistent with the numerical simulation.

\section{conclusion}
In conclusion, we realize a reconfigurable nanoelectromechanical network and demonstrate perfect coherent transfer in this on-chip device with different array networks.	
The system exhibits excellent reconfigurability to engineer various energy bands in the same on-chip device, which is crucial to the development of multifunctional and integrative nanomechanical metamaterials~\cite{Zheludev2016}.
Moreover, the electrically controllable NNN coupling breaks the hamper of spatial dimensions, which makes this platform more feasible to study pattern recognition~\cite{Vodenicarevic2017} and explore topology at high dimensions~\cite{Li2014,Maffei2018,An2018,Perez-Gonzalez2019}. Finally, the coupling method presented in this work shows outstanding tunability in oscillator networks, and this may also be employed in hybrid quantum computing architectures~\cite{Rabl2010}.

\begin{acknowledgements}
	
	This work was supported by the National Key R\&D Program of China (Grant No. 2018YFA0306600), the Chinese Academy of Sciences (Grants No. GJJSTD20170001 and No. QYZDY-SSW-SLH004), and Anhui Initiative in Quantum Information Technologies (Grant No. AHY050000). P. H. acknowledges support by the National Natural Science Foundation of China (Grant No. 11675163).
	
	T.T. and S.L. contributed equally to this work.
\end{acknowledgements}

\end{document}